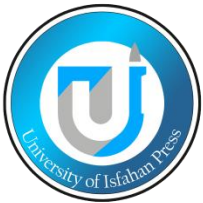

# Solving the Inverse Problem of Magnetic Induction Tomography Using Gauss-Newton Iterative Method and Zoning Technique to Reduce Unknown Coefficients

**Mohammad Reza Yousefi**[1,3], **Amin Dehghani**[2], **Ali Asghar Amini**[1,3], **S. M. Mehdi Mirtalaei**[1,3]

[1] Assistant Professor, Electrical Engineering Department, Najafabad Branch, Islamic Azad University, Iran

[2] Postdoctoral student, Department of Psychological and Brain Sciences, Dartmouth College, Hanover, NH, USA

[3] Smart Micro Grid Research Center, Najafabad Branch, Islamic Azad University, Iran

**Abstract:**
Magnetic Induction Tomography (MIT) is a promising modality for noninvasive imaging due to its contactless and nonionizing technology. In this imaging method, a primary magnetic field is applied by excitation coils to induce eddy currents in the material to be studied, and a secondary magnetic field is detected from these eddy currents using sensing coils. The image (spatial distribution of electrical conductivity) is then reconstructed using measurement data, the initial estimation of electrical conductivity, and the iterative solution of forward and inverse problems. The inverse problem can be solved using one-step linear, iterative nonlinear, and special methods. In general, the MIT inverse problem can be solved by Gauss-Newton iterative method with acceptable accuracy. In this paper, this algorithm is extended and the zoning technique is employed for the reduction of unknown coefficients. The simulation results obtained by the proposed method are compared with the real conductivity coefficients and the mean relative error rate is reduced to 24.22%. On the other hand, Gauss-Newton iterative method is extended for solving the inverse problem of the MIT, and sensitivity measurement matrices are extracted in different experimental and normalization conditions.

**مقاله پژوهشی**

# حل مسئلۀ معکوس مقطع‌نگاری القای مغناطیسی با استفاده از روش تکراری گوس - نیوتن و تکنیک ناحیه‌بندی برای کاهش ضرایب مجهول

محمدرضا یوسفی*[۱,۳]، امین دهقانی[۲]، علی اصغر امینی[۱,۳]، سید محمد مهدی میرطلایی[۱,۳]

۱- استادیار، دانشکده مهندسی برق، واحد نجف‌آباد، دانشگاه آزاد اسلامی، نجف‌آباد، ایران
mr-yousefi@iaun.ac.ir, mirtalaei.iaun@gmail.com

۲- دانشجوی پسادکتری، گروه علوم روانشناسی و مغز، کالج دارتموث، هانوفر، نیو، ایالات متحده
amin.dehghani@dartmouth.edu

۳- مرکز تحقیقات ریزشبکه‌های هوشمند، واحد نجف‌آباد، دانشگاه آزاد اسلامی، نجف‌آباد، ایران
amini33@yahoo.com

**چکیده:** در روش مقطع‌نگاری القای مغناطیسی با عبور یک جریان متناوب از یک یا چند سیم‌پیچ تحریک، یک میدان مغناطیسی تحریک در درون جسم مدنظر، ایجاد و ولتاژهای القایی در سیم‌پیچ‌های گیرنده اندازه‌گیری می‌شوند. بازسازی تصویر جسم با استفاده از نتایج حاصل از اندازه‌گیری، تخمین اولیه‌ای از ضرایب هدایت الکتریکی نواحی داخلی جسم، حل مسائل پیشرو و معکوس صورت می‌گیرد. مسئلۀ معکوس می‌تواند با استفاده از الگوریتم‌های مبتنی بر خطی‌سازی تکرارناپذیر، الگوریتم‌های تکراری غیرخطی و روش‌های خاص حل شود. یکی از معروف‌ترین این روش‌ها الگوریتم تکراری غیرخطی گوس – نیوتن است که نتایج خوبی را در مقطع‌نگاری امپدانس الکتریکی ارائه کرده است. در این مقاله این الگوریتم برای حل مسئلۀ معکوس مقطع‌نگاری القای مغناطیسی تعمیم یافته است. همچنین، با هدف کاهش تعداد مجهول‌ها از ایده ناحیه‌بندی استفاده شده است. برای ارزیابی کارایی روش پیشنهادی، نتایج حاصل با مقادیر واقعی مقایسه شدند که میانگین خطای نسبی مابین مقادیر رسانایی به‌دست‌آمده از حل مسئلۀ معکوس و مقادیر واقعی به مقدار ۲٤/۲۲ درصد کاهش پیدا کرده است. همچنین، الگوریتم تکراری غیرخطی گوس – نیوتن برای حل مسئلۀ معکوس مقطع‌نگاری القای مغناطیسی تعمیم و ماتریس حساسیت اندازه‌گیری‌ها در شرایط نرمالیزاسیون متفاوت استخراج و آزمایش شده‌اند.

**واژه‌های کلیدی:** الگوریتم‌های تکراری غیرخطی گوس – نیوتن، تصویربرداری القایی. حل مسئلۀ معکوس، مقطع‌نگاری القای مغناطیسی.

## ۱- مقدمه

مقطع‌نگاری الکتریکی یک روش تصویربرداری غیرتهاجمی از توزیع امپدانسی درون جسم با استفاده از انرژی‌دار کردن ناحیه مدنظر، انجام اندازه‌گیری‌های سطحی از نقاط متفاوت و بازسازی تصویر با به‌کارگیری نتایج حاصل از این اندازه‌گیری‌ها است. در مقطع‌نگاری، کمیت‌های فیزیکی مختلف مواد، امکان اندازه‌گیری‌های متفاوتی را ایجاد می‌کند که به پیدایش سیستم‌های مقطع‌نگاری متفاوتی منجر می‌شود. از این نمونه سیستم‌ها، مقطع‌نگاری امپدانس الکتریکی[۱] (EIT) [۱,۲]، مقطع‌نگاری ظرفیت خازنی الکتریکی[۲] (ECT) [۳,٤] و مقطع‌نگاری القای مغناطیسی[۳] (MIT) [٥,٦] هستند. در بین روش‌های

مختلف مقطع‌نگاری الکتریکی، مقطع‌نگاری القای مغناطیسی توجه محققان را امروزه به خود جلب کرده است. این روش برخلاف سایر روش‌های مقطع‌نگاری الکتریکی که از الکترودهای سطحی برای جمع‌آوری داده‌های اندازه‌گیری استفاده می‌کنند، براساس انجام اندازه‌گیری از روی سطح خارجی جسم و بدون تماس الکتریکی با آن عمل می‌کند. از مقطع‌نگاری الکتریکی در کاربردهای پزشکی و صنعتی در جاهایی که خواص هدایت الکتریکی و نفوذپذیری مغناطیسی اجسام اهمیت دارد، به‌ویژه پایش طولانی‌مدت بافت‌های بدن در کاربردهای بالینی، پایش فرآیندهای صنعتی، آزمون‌های غیر مخرب و ژئوفیزیک استفاده می‌شود [۷]. در زمینه سیستم‌های بالینی، این سیستم‌ها هنوز در فاز تحقیقاتی قرار دارند و تجاری نشده‌اند؛ اما در زمینه سیستم‌های صنعتی، این سیستم‌ها ازجمله سیستم معرفی‌شده در [۸] و [۹] به‌صورت تجاری در صنعت به‌کارگیری شده‌اند.

بازسازی تصویر در مقطع‌نگاری شامل دو مرحله حل مسئلهٔ پیشرو و مسئلهٔ معکوس است. مسئله پیشرو با پیداکردن یک سری خروجی یکتا با اعمال یک ورودی خاص و پارامترهای فیزیکی معلوم برای جسم هدف و با به‌کارگیری یک مدل مناسب حل تحلیلی یا عددی تعریف می‌شود. مسئله پیشرو در مقطع‌نگاری القای مغناطیسی شامل شبیه‌سازی میدان‌های حاصل از سیم‌پیچ‌های تحریک و جریان‌های گردابی جاری‌شده در درون جسم هدف و محاسبه ولتاژ اندازه‌گیری‌شده از سیم‌پیچ گیرنده به‌عنوان تابعی از ضرایب هدایت الکتریکی و مغناطیسی جسم هدف است. روش اجزای محدود، یکی از روش‌های مناسب در حل مسئلهٔ پیشرو مقطع‌نگاری القای مغناطیسی است [۱۰].

در نقطه مقابل، در مسئلهٔ معکوس، هدف ْ شناسایی پارامترهای فیزیکی سیستم با داشتن معلومات ورودی‌ها و خروجی‌های سیستم است. درواقع، بازسازی تصویر با استفاده از تخمین توزیع ضرایب‌های هدایت الکتریکی در یک سطح مقطع از جسم به‌وسیلهٔ اعمال یک میدان تحریک به جسم هدف و اندازه‌گیری‌های متعدد میدان ثانویه از سطح خارجی جسم صورت می‌گیرد [۱۲،۱۱]. به عبارت دیگر، مسئله شناسایی یک سیستم مجهول با معلوم‌بودن ورودی‌ها و خروجی‌های آن است. در حالت کلی به این نوع مسائل، مسئله معکوس گفته می‌شود. با حل مسئله پیشرو، جسم به‌صورت ریاضی مدل‌سازی می‌شود و سپس با اعمال ورودی‌ها (میدان تحریک) مطابق با آزمایشات واقعی به نقاط متناظر در مدل، مقادیر خروجی‌های ایجادشده در نقاط متناظر (ولتاژهای اندازه‌گیری‌شده از سیم‌پیچ‌های گیرنده) با مدل محاسبه می‌شوند. مقایسه داده‌های اندازه‌گیری واقعی و مدل‌سازی‌شده، مسئلهٔ معکوس را به سمت تخمین درست ضرایب فیزیکی بافت مدنظر هدایت می‌کند.

در سیستم‌های مقطع‌نگاری، القای مغناطیسی میدان تحریک به جسم، اعمال و میدان ثانویه نیز در قالب ولتاژ القاشده در سیم‌پیچ‌های گیرنده اندازه‌گیری می‌شود؛ بنابراین، تأثیر ضرایب رسانایی نواحی مرکزی جسم در شکل‌گیری ولتاژها بسیار کمتر از نواحی نزدیک‌تر به سطح جسم است. چنین مسائلی، مسائل بد وضع نامیده می‌شوند. به عبارت دیگر، اگر در مدل تغییرات عمده‌ای به وجود آید، با اعمال ورودی‌های یکسان، تغییرات خروجی جزئی خواهد بود؛ به‌ویژه اگر این تغییرات در نقاطی از مدل اعمال شوند که حساسیت خروجی نسبت به آن نقاط کم است [۱۳].

الگوریتم‌های بازسازی تصویر در مقطع‌نگاری القای مغناطیسی به طور عمده به سه دسته تقسیم می‌شوند: الگوریتم‌های مبتنی بر خطی‌سازی تکرارناپذیر، الگوریتم‌های تکراری غیرخطی و روش‌های خاص [۱۵،۱۴].

الگوریتم‌های خطی‌سازی با فرض اختلاف کم بین ضرایب رسانایی الکتریکی واقعی و ضرایب رسانایی الکتریکی مفروض، پایه‌گذاری می‌شوند. درمقابل، روش‌های غیرخطی تکراری در هر مرحله، ضرایب‌های مرحله قبل را با یک الگوریتم بهینه‌سازی تکراری تغییر می‌دهند؛ تا زمانی که خطا کمینه شود. در الگوریتم‌های مبتنی بر خطی‌سازی تکرارناپذیر، مبنای تفکر براساس مسائل کلاسیک فیزیکی و ریاضی است و سعی می‌شود با خطی‌سازی مسئله غیرخطی و با استفاده از روش‌های کلاسیک، مسئلهٔ معکوس حل شود. از این نوع روش‌ها روش پس‌افکنش[۴]، روش پس‌افکنش فیلترشده سریع[۵]، روش کوهن و بارسیه[۶] و روش

انحراف[7] هستند. تمامی این روش‌ها در ابتدا برای حل مسئلۀ معکوس مقطع‌نگاری الکتریکی به‌کارگیری شده‌اند [۱۶،۱۷]. پس از آن در مراجع [۱۸] و [۱۹] از روش نیوتن - گوس یک‌مرحله‌ای[8] که در دستۀ الگوریتم‌های تکرارناپذیر قرار می‌گیرد، با به‌کارگیری ۳۲ گرادیومتر[9] و ۱۶ سیم‌پیچ تحریک، برای بازسازی تصویر در مقطع‌نگاری القای مغناطیسی استفاده شده است. غیرخطی بودن ماهیت بازسازی تصویر در مقطع‌نگاری القای مغناطیسی باعث می‌شود دقت تصاویر به‌دست‌آمده از الگوریتم‌های یک‌مرحله‌ای مستقیم به دلیل خطی‌سازی کم باشد.

با حل مسئلۀ معکوس به شکل تکراری دقت تصاویر بازسازی‌شده بهبود می‌یابد؛ به همین دلیل، اغلب در سیستم‌های مقطع‌نگاری القای مغناطیسی روش‌های بازسازی تصویر تکرارشونده استفاده می‌شود. الگوریتم‌های تکراری براساس محاسبۀ ضریب‌های رسانایی الکتریکی جدید از آخرین توزیع ضریب‌های رسانایی الکتریکی و تولید تصویر تجدیدشده براساس تفاوت اندازه‌گیری‌های ولتاژ واقعی و اندازه‌گیری‌های شبیه‌سازی‌شده کار می‌کنند؛ البته باید توجه داشت روش‌های تکراری مستلزم صرف زمان محاسبۀ بسیار طولانی‌اند. الگوریتم غیرخطی تکراری معمول شامل روش‌های مونت کارلو[10]، نیوتن - رافسون[11] (نیوتن[12]) و گوس - نیوتن[13] شده (نیوتن - رافسون بهینه) است [۲۰،۲۱]. در مرجع [۲۲] الگوریتم غیرخطی تکراری مونت کارلو برای بازسازی تصویر مقطع‌نگاری القای مغناطیسی استفاده شده است. در این روش، زمان محاسبات به دلیل محاسبه ماتریس حساسیت با استفاده از روش تفاضل محدود بسیار طولانی است.

یکی دیگر از روش‌های غیرخطی تکراری برای حل مسئلۀ معکوس مقطع‌نگاری القای مغناطیسی، روش نیوتن - رافسون است. این روش حتی در نخستین تکرار، خطای کمتری نسبت برخی از روش‌های خطی یک‌مرحله‌ای از خود نشان می‌دهد. تحقیقات بعدی روی این روش، به ارائه روش نیوتن - رافسون بهینه‌شده یا روش گوس - نیوتن منجر شده است. روش گوس - نیوتن دارای دقت بیشتر و حساسیت کمتری نسبت به نویز در مقایسه با سایر الگوریتم‌های تکراری غیرخطی است [۲۳]. به دلیل نیازمندبودن الگوریتم‌های تکراری غیرخطی گوس - نیوتن به محاسبه ماتریس حساسیت برای تخمین پارامترهای مجهول، تا کنون روش‌های مختلفی برای این کار ارائه شده‌اند؛ ازجمله استفاده از روش اجزای محدود [۲۴] برای محاسبه ماتریس حساسیت. در [۲۵] از روشی مشابه برای محاسبه ماتریس حساسیت استفاده شده است؛ با این تفاوت که در محاسبات تنها حساسیت فاز اندازه‌گیری‌ها نسبت به ضریب رسانایی المان‌ها تأثیر داده شده است. به‌تازگی نیز یک مدل مداری برای محاسبه ماتریس حساسیت در [۲۶] پیشنهاد شده است. شاید بتوان این روش جدید محاسبه ماتریس حساسیت را گسترش‌یافتۀ مدل مداری ارائه‌شده در [۲۷] برای محاسبه ماتریس ژاکوبین در مقطع‌نگاری امپدانسی دانست.

در تحقیقات دیگری از روش‌های خاصی نظیر استفاده از شبکه‌های عصبی برای حل مسئلۀ معکوس و بازسازی تصویر مقطع‌نگاری القای مغناطیسی استفاده شده است [۲۸]. مشکل استفاده از شبکه‌های عصبی، نیازمندی این روش به داده‌های آموزش‌دهنده است؛ به این معنی که در این روش ابتدا باید با فرض یک جسم با ضریب‌های رسانایی الکتریکی معلوم شبکه آموزش پیدا کند و پس از آن، تغییر شکل‌های محدودی از جسم هدف قابل بازسازی است. در [۲۹] نیز از الگوریتمی خاص با به‌کارگیری تبدیل فوریه برای حل مسئلۀ معکوس استفاده شده است. محاسبات در این روش بسیار زمان‌بر است؛ تا جایی که به گفته ابداع‌کنندگان این روش، به دلیل زمان‌بر بودن محاسبات، روش برای مسائل سه‌بعدی قابل پیاده‌سازی و به‌کارگیری نیست.

در این مقاله این الگوریتم برای حل مسئلۀ معکوس مقطع‌نگاری القای مغناطیسی تعمیم و ماتریس حساسیت اندازه‌گیری‌ها، استخراج و بر یک مسئله دوبعدی به‌کارگیری خواهند شد. ساختار مقاله به این شرح است: در ابتدا در بخش ۲ روابط مورد نیاز برای حل مسئله پیشرو با استفاده از روش اجزای محدود و در بخش ۳ روابط مورد نیاز برای حل مسئلۀ معکوس با استفاده از روش تکراری غیر خطی گوس-نیوتن بیان شده‌اند. با هدف ارزیابی کارایی روش پیشنهادی در حل مسئلۀ معکوس، در بخش ۴ یک مسئله

دوبعدی در نظر گرفته شده است و نتایج حاصل از تخمین ضرایب با استفاده از روش پیشنهادی با نتایج واقعی مقایسه شده‌اند. درنهایت، نتیجه‌گیری تحقیق در بخش ۵ بیان شده است.

## ۲- حل مسئلۀ پیشرو

در حل مسئلۀ پیشرو مقطع‌نگاری القای مغناطیسی، محیط مورد حل به‌طور معمول با یک معادله دیفرانسیل با مشتقات جزئی از نوع هلمهلتز به فرم عمومی زیر مدل‌سازی می‌شود [۳۰،۳۱]:

$$-\nabla.(\alpha(x,y)\nabla u(x,y)) + \beta(x,y)u(x,y) = f(x,y) \quad (1)$$

که $\alpha(x,y)$ و $\beta(x,y)$ پارامترهای فیزیکی محیط و $f(x,y)$ تابع تحریک هر سه تابعی از مکان‌اند. برای حل این معادله، از هر یک از دو روش بهینه‌سازی ریتز یا گالرکین برای فرمول‌بندی روش اجزای محدود بهره برده می‌شود [۳۲،۳۳]. برای این منظور، ابتدا ناحیه اجزای محدود با به‌کارگیری تعدادی المان تقسیم‌بندی می‌شود، سپس برای هر المان عددی منحصربه‌فرد اختصاص می‌یابد. گره‌های هر المان نیز به‌صورت محلی[14] شماره‌گذاری می‌شوند. شماره‌گذاری منحصربه‌فرد برای هر گره در کل دامنه، شماره‌گذاری سراسری[15] نامیده می‌شود. تغییرات متغیر مجهول، در طول یک المان با انتخاب مناسب توابع پایه، تقریب زده می‌شود. در روش اجزای محدود، معمولاً از توابع چندجمله‌ای به‌عنوان توابع پایه استفاده می‌شود؛ زیرا مشتق و انتگرال این‌گونه توابع در مقایسه با توابع دیگر، به‌آسانی محاسبه‌پذیر است. در حالت دوبعدی با فرض المان‌های مثلثی برای المان‌بندی ناحیه، تغییرات متغیر مجهول $u(x,y)$ در داخل المان e-ام با یک تابع خطی به شکل زیر تقریب زده می‌شود [۳۴]:

$$f_i = C_0 + C_1 + C_2^2 \quad (2)$$

$$u^e(x,y) = a_1^e + a_2^e x + a_3^e y \quad (3)$$

با محاسبۀ ضرایب مجهول $a_1^e$, $a_2^e$ و $a_3^e$ بر حسب مقدار پتانسیل مجهول بر گره‌های هر المان و جایگذاری آنها در بسط (۲)، تغییرات متغیر مجهول $u^e(x,y)$ در داخل هر المان با رابطه (۴) بیان می‌شود:

$$u^e(x,y) = \sum_{j=1}^{3} N_j^e(x,y) u_j^e \quad (4)$$

در این رابطه $u_j^e$ مقدار پتانسیل مجهول روی گره j‌ام المان e و $N_j^e$ تابع پایه درون‌یاب متناظر با این گره است. به این ترتیب و با انتخاب این توابع پایه به‌عنوان وزن باقی‌مانده معادله بیان‌شده در رابطۀ (۱)، باقی‌مانده وزن‌دارشده در هر المان با فرض ثابت‌بودن ثابت‌های فیزیکی در آن المان، به شکل ماتریسی زیر دردسترس است:

$$\{R^e\} = [K^e]\{u^e\} - \{F^e\} - \{g^e\} \quad (5)$$

که در آن:

$$k_{ij}^e = \iint_{\Omega^e} \left\{ \alpha^e \frac{\partial N_i^e(x,y)}{\partial x} \frac{\partial N_j^e(x,y)}{\partial x} + \alpha^e \frac{\partial N_i^e(x,y)}{\partial y} \frac{\partial N_j^e(x,y)}{\partial y} + \beta^e N_i^e(x,y) N_j^e(x,y) \right\} dxdy \quad for\ i,j = 1,2,3 \quad (6)$$

$$F_i^e = \iint_{\Omega^e} N_i^e f^e dxdy \quad for\ i = 1,2,3 \quad (7)$$

$$g_i^e = \oint_{\Gamma^e} N_i^e(x,y) D.\hat{n}^e d\Gamma \quad for\ i = 1,2,3 \quad (8)$$

$$\{u^e\} = \{u_1^e \quad u_2^e \quad u_3^e\}^T \quad (9)$$

با گسترش این رابطه روی تمامی N المان تشکیل‌دهندۀ ناحیه حل، رابطه ماتریسی زیر به دست آمد:

$$[K][u] + [g] = [F] \quad (10)$$

در این رابطه K و F به‌ترتیب ماتریس سختی و بردار تحریک هستند. u نیز بیان‌کنندۀ تغییرات $u(x,y)$ بر گره‌های ناحیه است. با گسترش این دستگاه روی تمام المان‌ها و اعمال شرایط مرزی مطابق روش ارائه‌شده در [۳۵] درنهایت، دستگاه معادلات بیان‌شده به‌صورت زیر ساده‌سازی می‌شود:

$$\begin{bmatrix} K & H_v \\ H_u & I \end{bmatrix} \begin{bmatrix} u \\ \lambda \end{bmatrix} = \begin{bmatrix} F \\ q \end{bmatrix} \quad (11)$$

در این رابطه $H_v$، $H_u$ و $I$ برای اعمال شرایط مرزی رابین[16] استفاده می‌شوند. برای اعمال شرایط مرزی ترکیبی رابین (حالت کلی در برگیرندهٔ شرایط مرزی نیومن[17] و دیریشله[18]) برای هر نقطه $P_k$ روی مرز خواهیم داشت:

$$\alpha_k \frac{\partial u(x,y)}{\partial \hat{n}} \cdot \hat{n} \bigg|_{P_k} + \gamma_k u(P_k) = \sum_{j=1}^{M} c_j (\alpha_k \frac{\partial \varphi_j(x,y)}{\partial \hat{n}} \cdot \hat{n} \bigg|_{P_k} + \gamma_k \varphi_j(P_k)) = q_k \quad (12)$$

که در آن $\hat{n}$ بردار نرمال بر مرز و $\frac{\partial u(x,y)}{\partial \hat{n}} \cdot \hat{n} \bigg|_{P_k}$ معرف مقدار تغییرات تابع u در راستای این بردار در نقطه $P_k$ روی مرز است. مقادیر $\alpha_k$، $\gamma_k$ و $q_k$ با توجه به ماهیت مسئله تعیین شده‌اند و مقادیری معلوم‌اند [۳۶].

درایه‌های ماتریس $H_v$ روی قسمت‌های مرزی رابین مقدار می‌گیرد و برای مرزهای نیومن همگن و دیریشله مقدار صفر دارد. در هر سطر از این ماتریس به‌ازای گره i قرارگرفته روی مرز، برای دو قسمت متصل به آن عدد ۰/۵ و برای سایر قسمت‌های مرزی عدد صفر در نظر گرفته می‌شود:

$$H_v = \frac{1}{2} \begin{bmatrix} 0 & \cdots & & & 0 \\ & \ddots & & & \\ \vdots & & 1|_{i,s} & 1|_{i,s+1} & \vdots \\ & & & \ddots & \\ 0 & \cdots & & & 0 \end{bmatrix}_{n \times BS} \quad (13)$$

در این رابطه BS تعداد قسمت‌های مرز و n تعداد گره‌های تعریف‌شده در ناحیه حل هستند. همچنین، با در نظر گرفتن شرط مرزی رابین، برای قسمت مرزی sام، در سطر s-ام زیرماتریس $H_u$، ستون متناظر با گره ابتدایی قسمت s-ام مقدار $\frac{\gamma^{(s)}}{2}$ و سایر ستون‌ها مقدار صفر خواهند داشت:

$$H_u = \frac{1}{2} \begin{bmatrix} 0 & \cdots & \gamma^1|_{1,i^1} & \cdots & 0 \\ \vdots & & \ddots & & \vdots \\ 0 & \cdots & \gamma^{BS}|_{BS,i^{BS}} & \cdots & 0 \end{bmatrix}_{BS \times n} \quad (14)$$

ماتریس I نیز یک ماتریس قطری است که روی قطر آن، مقدار 1/ls- قرار می‌گیرد:

$$I = \begin{bmatrix} -\frac{1}{l^1} & \cdots & 0 \\ \vdots & \ddots & \vdots \\ 0 & \cdots & -\frac{1}{l^{BS}} \end{bmatrix}_{BS \times BS} \quad (15)$$

همچنین، بردار q نیز تعیین‌کنندهٔ مقدار شرط مرزی و بردار مجهول $\lambda$ بیان‌کنندهٔ ضریب‌های لاگرانژ در BS قسمت مرزی است:

$$q = \begin{bmatrix} q^1 & \cdots & q^{BS} \end{bmatrix}^T \quad (16)$$

$$\lambda = \begin{bmatrix} \lambda^1 & \cdots & \lambda^{BS} \end{bmatrix}^T \quad (17)$$

حال با توجه به اینکه خروجی‌های مسئله پیشرو ولتاژهای القایی در سیم‌پیچ‌های گیرنده‌اند، این ولتاژها باید شبیه‌سازی شوند. محاسبه این ولتاژها با بهره‌گیری از قانون القای فاراده به شکل زیر امکان‌پذیر است [۳۷]:

$$V_{ind} = \oint_c \vec{E} \cdot dl \quad (18)$$

در این رابطه $\vec{E}$ شدت میدان الکتریکی عبوری از داخل سیم‌پیچ و c مسیر بسته‌شده با سیم‌پیچ است. حال با استفاده از رابطه $\vec{E} = -j\omega \vec{A} + \vec{E}_s$ که $\vec{E}_s$ شدت میدان الکتریکی ایجادشده با منبع تحریک در صورت وجود منبع میدان الکتریکی در محیط مدنظر است و با صرف نظر کردن از $\vec{E}_s$ در سیم‌پیچ گیرنده، خواهیم داشت:

$$V_{ind} = -j\omega \oint_c A_z \cdot dl \quad (19)$$

بنابراین، اختلاف ولتاژ ظاهرشده در دو سر سیم‌پیچ گیرنده با رابطه زیر محاسبه می‌شود:

$$V_m = j\omega L(A_1 - A_2) \quad (20)$$

در این رابطه، $A_1$ و $A_2$ اندازه بردار پتانسیل در دو هادی رفت و برگشت سیم‌پیچ در سطح مقطع مدنظر است که می‌تواند با مقادیر بردار پتانسیل مغناطیسی در وسط هادی‌ها جایگزین شوند.

## ۳- حل مسئلهٔ معکوس

بعد از تعریف مسئله معکوس، در این قسمت الگوریتمی برای حل آن برای دستیابی به مقادیر مجهول رسانایی المان‌ها از روی ولتاژهای اندازه‌گیری ارائه می‌شود. حل مسئله در دو مرحله انجام می‌شود؛ مرحله نخست، انتخاب تابع هزینه و مرحله دوم، تعیین پارامترهای مجهول با استفاده از یک روش بهینه‌سازی است.

### ۳-۱- انتخاب تابع هزینه

تابع هزینه برای حل مسئلهٔ معکوس مقطع‌نگاری مغناطیسی، تابع هزینه مجموع مربعات خطای وزن‌دار شده یا به‌اختصار LS انتخاب می‌شود. این تابع هزینه در حالت کلی به‌صورت زیر تعریف می‌شود:

$$J(p) = e^T(p).Q.e(p) \tag{۲۱}$$

که در آن، p ماتریس ستونی پارامترهای مجهول، e ماتریس ستونی خطاها و Q تابع وزنی است. تابع خطا e به فرم زیر تعریف می‌شود:

$$e(p) = V^m - f(p) \tag{۲۲}$$

در این رابطه، $V^m$ ماتریس ستونی ولتاژهای اندازه‌گیری‌شده از سیم‌پیچ‌ها است که در صورت نبود داده‌های واقعی، این اندازه‌گیری‌ها به شرحی که در مسئله پیشرو گفته شد، از حل مسئله پیشرو با مش ریز شبیه‌سازی می‌شوند. تابع وزنی Q نیز در حالت کلی به صورت ماتریس مربعی به شکل زیر تعریف می‌شود:

$$Q = [w_{ij}] \tag{۲۳}$$

ساده‌ترین انتخاب Q=I/2 است که I ماتریس یکه است. بدین ترتیب تابع هزینه به شکل زیر تعریف خواهد شد.

$$J(p) = \frac{1}{2} e^T(p). e(p) \\ = \frac{1}{2}(V^m - f(p))^T.(V^m - f(p)) \tag{۲۴}$$

در این رابطه $V^m$ ولتاژهای اندازه‌گیری‌شده از سیستم واقعی و $f(p)$ خروجی مسئله پیشرو با فرض ماتریس ستونی پارامترهای اولیه $p$ است.

### ۳-۲- روش بهینه‌سازی گوس - نیوتن

در روش بهینه‌سازی گوس - نیوتن برای تخمین بردار پارامترهای مجهول $p$، از بسط تیلور به شکل زیر استفاده می‌شود:

$$J(p^{k+1}) = J(p^k + \Delta p) = J(p^k) + \\ \left[\frac{\partial J(p)}{\partial p}\bigg|_{p^k}\right]^T . \Delta p + \frac{1}{2}\left[\Delta p. \frac{\partial^2 J(p)}{\partial^2 p}\bigg|_{p^k}\right]^T . \Delta p \\ + o\|\Delta p\|^3 \tag{۲۵}$$

با جایگذاری رابطهٔ تابع هزینه در این رابطه و صرف‌نظرکردن از جزء $f''(p^k)$، رابطهٔ تکراری به‌روزرسانی پارامترهای مجهول در این روش با رابطهٔ زیر انجام می‌شود:

$$p_{e\times 1}^{k+1} = p_{e\times 1}^k + \Delta p_{e\times 1}^k \\ = p_{e\times 1}^k - \left[\left(f'(p^k)\right)^T . f'(p^k)\right]_{e\times e}^{-1} \\ . \left(f'(p^k)\right)_{e\times m}^T . \left(V^m - f(p^k)\right)_{m\times 1} \tag{۲۶}$$

در این رابطه، جزء $f'(p^k)$ بیان‌کنندهٔ تغییرات اندازه‌گیری ولتاژ از سیم‌پیچ‌ها نسبت به تغییرات پارامترها (رسانایی هر المان) در هر مرحله از تصحیح پارامترها است که به ماتریس حساسیت معروف است. $e$ تعداد پارامترهای مجهول (تعداد المان‌ها در روش المان محدود) و $m$ تعداد اندازه‌گیری‌های ممکن است.

شکل (۱) روند نمای جستجو الگوریتم گوس - نیوتن را نشان می‌دهد. الگوریتم از یک حدس اولیه برای ضرایب رسانایی مجهول آغاز می‌شود و با اعمال این ضرایب به مسئله پیشرو، ولتاژهای شبیه‌سازی‌شده ناشی از این ضرایبِ تخمین زده شده به دست می‌آیند. پس از آن خطای بین این ولتاژهای شبیه‌سازی‌شده خروجی مسئله پیشرو با ولتاژهای واقعی، ضرایب تخمین زده شده را تصحیح می‌کنند. الگوریتم تا کمترشدن خطای بین ولتاژهای اندازه‌گیری‌شده و خروجی مسئله پیشرو از معیار خطای ε ادامه می‌یابد. $k$ شمارندهٔ این الگوریتم تکراری است.

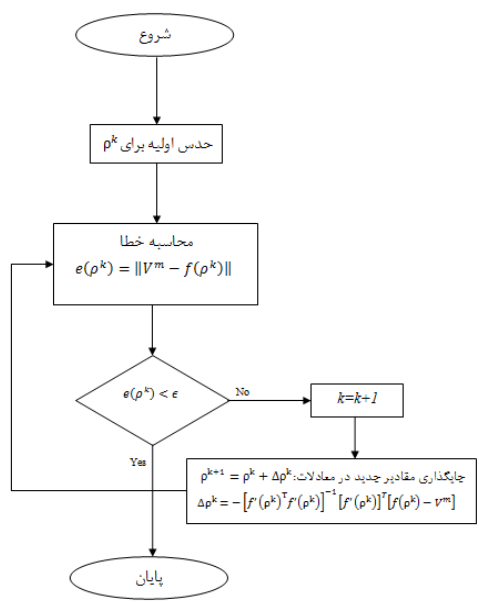

**شکل (۱): فلوچارت الگوریتم گوس – نیوتن.**

### ۳-۳- نتایج حل مسئلۀ معکوس

شکل (۲) شمایی از یک سیستم مقطع‌نگار القای مغناطیسی ۸ سیم‌پیچه را نشان می‌دهد که هم نقش سیم‌پیچ تحریک و هم نقش سیم‌پیچ گیرنده را بازی می‌کنند. با تولید مش سیستم مقطع‌نگاری القای مغناطیسی مفروض، ناحیه هدف به ۴۱۶ المان مثلثی تقسیم می‌شود. کل ناحیه مورد حل برای حل مسئلۀ پیشرو به روش اجزای محدود به ۱۵۲۸ المان مثلثی تقسیم شده است. با توجه به اینکه تعداد اندازه‌گیری‌های ممکن در این حالت ۲۸ مورد خواهد بود، عملاً تخمین ضریب‌های رسانایی ۴۱۶ المان قرارگرفته در ناحیه هدف غیرممکن است. همچنین، کاهش تعداد المان‌ها دقت حل مسئله پیشرو را به‌شدت کاهش خواهد داد. برای رفع این مشکل، ناحیه هدف به ۲۸ زیر ناحیه، تقسیم می‌شود و ضریب‌های رسانایی المان‌های قرارگرفته در هر زیرناحیه با یکدیگر یکسان فرض می‌شوند و با یکدیگر تغییر می‌کنند. با استفاده از این روش، عملاً تعداد ضریب‌های رسانایی مجهول به ۲۸ مورد کاهش می‌یابد. این ۲۸ زیرناحیه در شکل (۳) با رنگ قهوه‌ای، سیم‌پیچ‌ها با رنگ قرمز و هوا با رنگ خاکستری متمایز شده‌اند. همچنین، در این شبیه‌سازی فرض شده است که دو میله با ضریب رسانایی صد برابر زمینه ($\sigma = 100 S/m$) در کناره‌های ناحیه هدف قرار گرفته‌اند که در این شکل این میله‌ها نیز با رنگ نارنجی مشخص شده‌اند.

در این شبیه‌سازی، ضریب رسانایی هوا صفر، سیم‌پیچ‌ها $\sigma = 5.8 \times 10^7 S/m$، ناحیه هدف $\sigma = 1 S/m$ و میله‌ها $\sigma = 100 S/m$ فرض شده‌اند. ضریب‌های نفوذپذیری نسبی مغناطیسی هوا، سیم‌پیچ‌ها و شیء هدف $\mu_r$ برابر یک فرض شده‌اند. در این شبیه‌سازی برای تأثیردادن تفاوت امپدانسی سیم‌پیچ‌ها، فرض شد چگالی جریان‌های ۱۰۰۰، ۱۱۰۰، ۹۰۰، ۱۰۰۰، ۱۲۰۰، ۱۰۰۰، ۸۰۰ و ۱۰۰۰ کیلوآمپر بر متر به‌ترتیب به سیم‌پیچ‌های ۱ تا ۸ اعمال شده‌اند.

سطح مقطع سیم‌پیچ‌ها ۰/۵ سانتیمتر مربع و فاصله هوایی مابین سیم‌پیچ‌ها و ناحیه هدف برابر ۰/۵ سانتیمتر فرض شده است. ناحیه هدف نیز به‌صورت یک دایره با قطر ۱۴ سانتیمتر فرض شده است. شکل (۴) مش اولیه کل ناحیه مدنظر با ۱۵۲۸ المان مثلثی را برای حل مسئله پیشرو به روش اجزای محدود نمایش می‌دهد. روش استفاده‌شده برای تصحیح ضریب‌های رسانایی روش غیرخطی تکراری گوس – نیوتن با استفاده از رابطه تکراری به‌روزرسانی پارامترهای مجهول است. روند تصحیح پارامترها از مقدار ضریب رسانایی پس‌زمینۀ ناحیه هدف $\sigma = 1 S/m$ آغاز می‌شود. به دلیل بسیار کوچک‌بودن مقادیر ولتاژهای اندازه‌گیری‌شده، هر کدام از اندازه‌گیری‌ها به همراه سطر متناظر آن از ماتریس حساسیت، نسبت به مقادیر شبیه‌سازی‌شده ولتاژ اندازه‌گیری‌شده از سیم‌پیچ‌ها بدون حضور جسم هدف نرمالیزه می‌شوند. پس از آن، خطای مابین ولتاژهای اندازه‌گیری‌شدۀ خروجی مسئله پیشرو با رسانایی مفروض و ولتاژهای اندازه‌گیری‌شدۀ حاصل شبیه‌سازی در حضور مقادیر رسانایی واقعی، محاسبه و وارد الگوریتم بازگشتی با هدف تصحیح رسانایی المان‌ها می‌شوند. این روند تا برقراری یکی از شرایط توقف شامل بیشترشدن تعداد مراحل تصحیح از محدودیت اعمال‌شده (۵۰ تکرار در این مثال)، کمترشدن خطای مابین ولتاژهای اندازه‌گیری خروجی مسئله پیشرو با ولتاژهای اندازه‌گیری‌شدۀ واقعی (حاصل شبیه‌سازی در حضور مقادیر رسانایی واقعی) از محدودیت اعمال‌شده (۰/۰۱ در این

مثال) یا کمتر‌شدن تغییرات ضـریب‌هـای رسانایی از محدودیت اعمال‌شده (۰/۱ در این مثال) ادامه می‌یابد. شرط توقف در روش غیرخطی تکراری گوس – نیـوتن برقراری شرط میانگین خطای بین ولتاژهای اندازه‌گیری‌شدهٔ واقعی (حاصل شبیه‌سازی در حضـور مقـادیر رسانایی واقعـی) و شبیه‌سـازی‌شـده (حاصـل شبیه‌سازی در حضور مقـادیر رسانایی تخمین زده شده) کمتر از ۰/۰۱ ولـت تنظیـم شـده است. بدین ترتیب پس از انجـام ۳۷ مرحلـه تکـرار، رونـد تصحیح پارامترهای مجهول با برقراری ایـن شـرط متوقـف شده است. در این حالت میانگین خطای نسبی مابین مقادیر رسانایی به‌دست‌آمده از حل مسئلۀ معکوس و مقادیر واقعی قرارگرفته در محیط نسبت به مقدار بیشینۀ رسانایی ناحیـه هدف به ۲٤/۲۲ درصد بـه دسـت آمـده اسـت. شـکل (۵) تغییرات تابع هزینه بر حسب مراحل مختلف تکرار در مثال دو میله با نرمالیزاسیون چندگانـه و حـل مسـئلۀ پیشـرو بـه روش اجزای محدود را نشان می‌دهد. تصویر بازسازی‌شـده در این شبیه‌سازی در شکل (۶) نمایش داده شده است.

با هدف بررسی همگرایی خطا در روش پیشنهادی، مش ناحیه تغییر کرده و با بازسازی مش، مـش جدیـد بـا تعـداد ۶۱۱۲ المان مثلثی روی کل ناحیه حـل ایجـاد شـده اسـت. شکل (۷) تصویر این مش بازسازی‌شده را نشان مـی‌دهـد. پس از حـل مسـئلۀ معکـوس بـا بـه‌کـارگیری ایـن مـش بازسازی‌شده در حل مسئله پیشرو، بـا گذشـت ٤۰ مرحلـه تکرار الگوریتم متوقف می‌شود که شکل (۸) تغییـرات تـابع هزینه بر حسب مراحل مختلف تکرار با بـه‌کـارگیری مـش بازسازی‌شـده در حـل مسـئلۀ پیشـرو را نشـان مـی‌دهـد. همگرایی روش حل مسئلهٔ معکوس در این حالت نیز کاملاً مشهود است. شکل (۹) نیز تصویر بازسازی‌شده در ایـن حالت را نمایش می‌دهد. تغییرات شکل بازسازی‌شده نسبت به حالت قبل ناشی از افزایش ابعاد مـاتریس حساسـیت بـه سبب افزایش تعداد المان‌هـا و بـه تبـع آن، افـزایش درجـۀ آزادی در مسئلۀ معکـوس اسـت. در ایـن حالـت، میـانگین خطای نسبی مابین مقادیر رسانایی بـه‌دسـت‌آمـده از حـل مسئلۀ معکوس و مقادیر واقعی قرارگرفته در محـیط نسـبت به مقدار بیشینۀ رسانایی ناحیـه هـدف بـه ۱۶/۶٤ درصـد می‌رسد که کاهش میانگین خطای نسبی در حالت استفاده از

مش بازسازی‌شده در حل مسئله پیشرو را نشان می‌دهد.

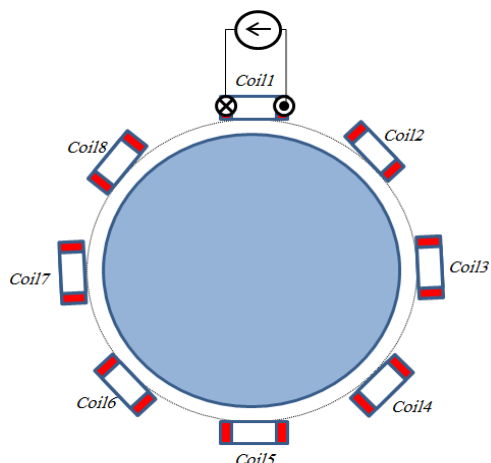

شکل (۲): سیستم مقطع‌نگار القای مغناطیسی ۸ سیم‌پیچه

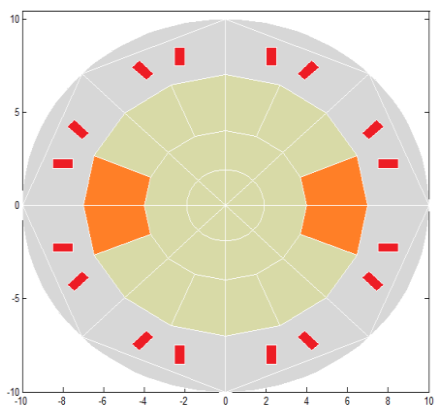

شکل (۳): تقسیم‌بندی ناحیه هدف به ۲۸ زیرناحیه و حضور دو میله با ضریب رسانایی ۱۰۰ برابر پس‌زمینه در کناره‌های ناحیه هدف

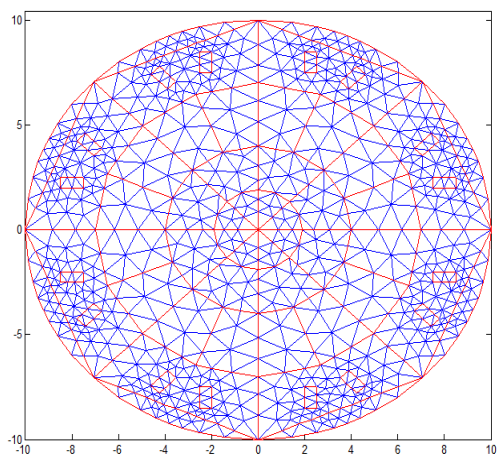

شکل (٤): مش اولیه کل ناحیه حل

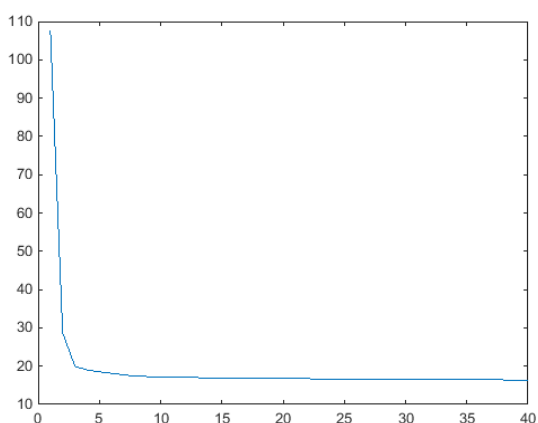

شکل (۸): تغییرات تابع هزینه در مراحل مختلف تکرار (مش بازسازی‌شدهٔ مسئله پیشرو)

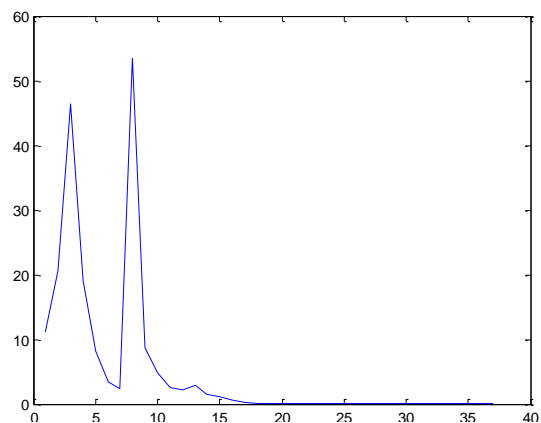

شکل (۵): تغییرات تابع هزینه در مراحل مختلف تکرار (مش اولیه مسئله پیشرو)

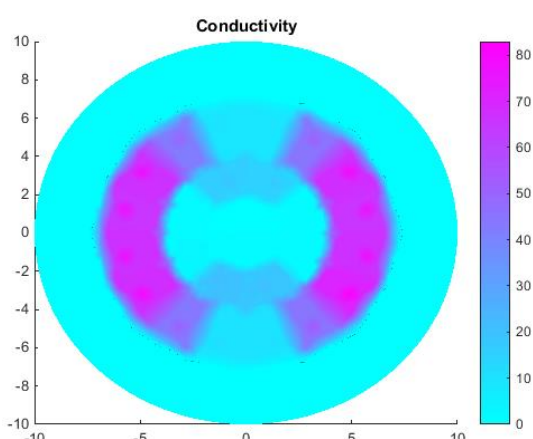

شکل (۹): تصویر بازسازی‌شده با روش پیشنهادی (مش بازسازی‌شده مسئله پیشرو)

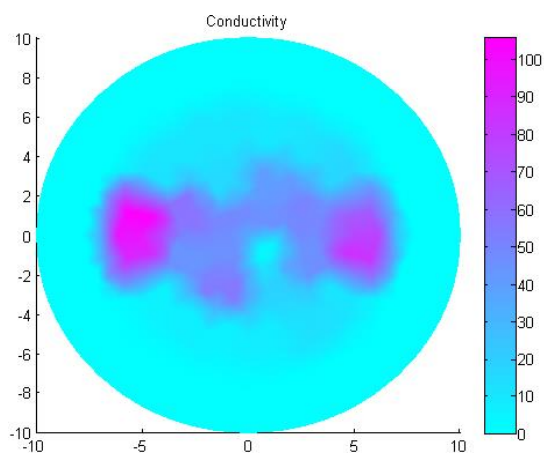

شکل (۶): تصویر بازسازی‌شده با روش پیشنهادی (مش اولیه مسئله پیشرو)

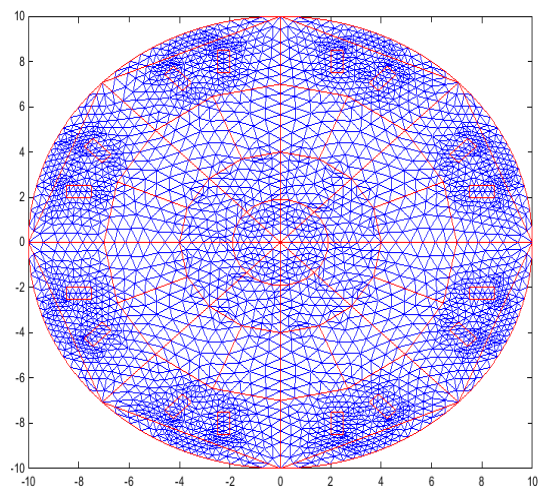

شکل (۷): مش بازسازی‌شدهٔ کل ناحیه حل

## ۴- نتیجه‌گیری

مقطع‌نگاری القای مغناطیسی یک روش تصویربرداری غیرهجومی و غیرتداخلی از داخل یک جسم هدف، براساس انجام اندازه‌گیری از روی سطح خارجی جسم و بدون تماس الکتریکی با آن است. در این مقاله، الگوریتم تکراری غیرخطی گوس – نیوتن برای حل مسئلۀ معکوس مقطع‌نگاری القای مغناطیسی تعمیم و ماتریس حساسیت اندازه‌گیری‌ها در شرایط نرمالیزاسیون متفاوت، استخراج و بر روی یک مسئله تست دوبعدی آزمایش شدند. برای بازسازی تصویر به دلیل بسیار کوچک‌بودن ولتاژهای اندازه‌گیری‌شده و به تبع آن، خطای بین مسئله پیشرو و نتایج واقعی، تغییرات رسانایی بسیار شدیدند و باعث

ناپایــداری حــل مــی‌شــوند. بــرای رفــع ایــن مشــکل، از نرمالیزه‌کردن داده‌ها نسبت به یک مبنای خــاص بهــره بــرده مــی‌شــود. بــومی‌ســازی روش‌هــای حــل مســئلهٔ معکــوس مقطع‌نگاری، باعث تعمیم آن بر انواع سیستم‌هــا و محاســبه ماتریس حساسیت متناسب با شرایط مسئله می‌شود.

منظور از بهینه‌سازی مصــرف انــرژی انتخــاب الگوهــا، اتخاذ و به‌کــارگیری روش‌هــا و سیاســت‌هایی در مصــرف انرژی الکتریکی است. ساختمان‌های مسکونی بخش مهمی از مصرف‌کنندگان انرژی الکتریکی به شمار مــی‌آینــد. ورود تکنولــوژی سیســتم مــدیریت هوشــمند بــه ســاختمان‌هــای مسکونی، تا حدودی مصرف انرژی الکتریکی را بهینه کــرده است.

# مراجع

---

[1] Electrical impedance tomography

[2] Electrical capacitance tomography

[3] Magnetic induction tomography

[4] Back-projection

[5] Fast filtered back projection method

[6] Cohen and Barcie

[7] Perturbation method

[8] Gauss–Newton one step method

[9] Gradiometer

[10] Monte Carlo

[11] Newton-Raphson

[12] Newton method
[13] Gauss-Newton method
[14] Local
[15] Globa
[16] Robin
[17] Neuman
[18] Dirishlet